\documentclass{EPS}

\usepackage[subrefformat=parens]{subcaption}
\usepackage{graphicx}
\usepackage{lineno}

\usepackage{color}

\title{Operational Solar Flare Prediction Model Using Deep Flare Net}
\author{
Naoto Nishizuka, Applied Electromagnetic Research Institute,
National Institute of Information and Communications Technology, Tokyo 184-0015, Japan, nishizuka.naoto@nict.go.jp\\
Y\^uki Kubo, Applied Electromagnetic Research Institute, National Institute of Information and Communications Technology, Japan, kubo@nict.go.jp\\
Komei Sugiura, Department of Information and Computer Science, Keio University, Japan, komei.sugiura@keio.jp\\
Mitsue Den, Applied Electromagnetic Research Institute, National Institue of Information and Communications Technology, Japan, den@nict.go.jp\\
Mamoru Ishii, Applied Electromagnetic Research Institute, National Institute of Information and Communications Technology, Japan, mishii@nict.go.jp
}

\abstract{
We developed an operational solar flare prediction model using deep neural networks, named Deep Flare Net (DeFN). DeFN can issue probabilistic 
forecasts of solar flares in two categories, such as $\geq$M-class and $<$M-class events or $\geq$C-class and $<$C-class events, occurring in 
the next 24 h after observations and the maximum class of flares occurring in the next 24 h. DeFN is set to run every 6 h and has been operated 
since January 2019. The input database of solar observation images taken by the Solar Dynamic Observatory (SDO) is downloaded from the data 
archive operated by the Joint Science Operations Center (JSOC) of Stanford University. Active regions are automatically detected from 
magnetograms, and 79 features are extracted from each region nearly in real-time using multiwavelength observation data. Flare labels are 
attached to the feature database, then the database is standardized and input into DeFN for prediction. DeFN was pretrained using the datasets 
obtained from 2010 to 2015. The model was evaluated with the skill score of the true skill statistics (TSS) and achieved predictions with TSS = 0.80 
for $\geq$M-class flares and TSS = 0.63 for $\geq$C-class flares. For comparison, we evaluated the operationally forecast results from January 
2019 to June 2020. We found that operational DeFN forecasts achieved TSS = 0.70 (0.84) for $\geq$C-class flares with the probability threshold 
of 50 (40) \%, although there were very few M-class flares during this period and we should continue monitoring the results for a longer time. Here, 
we adopted a chronological split to divide the database into two for training and testing. The chronological split appears suitable for evaluating 
operational models. Furthermore, we proposed the use of time-series cross-validation. The procedure achieved TSS = 0.70 for $\geq$M-class 
flares and 0.59 for $\geq$C-class flares using the datasets obtained from 2010 to 2017. Finally, we discuss the standard evaluation methods for 
operational forecasting models, such as the preparation of observation, training, and testing datasets, and the selection of verification metrics.
}


\begin{document}

\maketitle

\section{1.	Introduction}

The mechanism of solar flares is a long-standing puzzle. Solar flares emit X-rays, highly energetic particles, and coronal mass ejections (CMEs) into 
the interplanetary space in the heliosphere, whereby these flares become one of the origins of space weather phenomena \citep[e.g.,][]{sch05, fle11, 
liu14, mos15}. The prediction of flares is essential for reducing the damage to technological infrastructures on Earth. Solar flares are triggered by a 
newly emerging magnetic flux or magnetohydrodynamic instability to release excess magnetic energy stored in the solar atmosphere \cite[e.g.,][]{aul10, 
shi11, che14, wanh17, tor19}. Such phenomena are monitored by the Solar Dynamic Observatory \cite[SDO;][]{pes12} and the Geostationary Orbital 
Environment Satellite (GOES), and observation data are used for the prediction of flares.\\

Currently, flare prediction is tackled by the following four approaches: (i) empirical human forecasting \citep{cro12, dev14, kub17, mur17}, (ii) statistical 
prediction methods \citep{lee12, blo12, mcc16, lek18}, (iii) machine learning methods \citep[e.g.,][and references therein]{bob15, mur15, nis17}, and 
(iv) numerical simulations based on physics equations \citep[e.g.,][]{kus12, kus20, ino18, kor20}. Some of the models have been made available for 
community use at the Community Coordinated Modeling Center (CCMC) of NASA \citep[e.g.,][]{gal02, shi03, col08, col09, kri09, ste11, fal11, fal12}. 
It is useful to show the robust performance of each model, and in benchmark workshops, prediction models were evaluated for comparison, where 
methods that included machine learning algorithms as part of their system were also discussed \citep{bar16, lek19, par20}.\\

Recently, the application of supervised machine learning methods, especially deep neural networks (DNNs), to solar flare prediction has been a hot 
topic, and their successful application in research has been reported \citep{hua18, nis18, par18, che19, dom19, liu19, zhe19, bha20, jia20, li20, pan20, 
yi20}. However, there is insufficient discussion on how to develop the methods available to real-time operations in space weather forecasting offices, 
including the methods for validation and verification of the models. Currently, new physical and geometrical (topological) features are applied to flare 
prediction using machine learning \citep[e.g.,][]{wanj20, des20}, and it has been noted that training sets may be sensitive to which period in the solar 
cycle they are drawn from. \citep{wanx20}.\\

It has been one year since we started operating our flare prediction model using DNNs, which we named Deep Flare Net \citep[DeFN:][]{nis18}. 
Here, we evaluate the prediction results during real-time operations at the NICT space weather forecasting office in Tokyo, Japan. In this paper, 
we introduce the operational version of DeFN in sections 2 and 3, and we show the prediction results in section 4. We propose the use of time-series 
cross-validation (CV) to evaluate operational models in section 5. We summarize our results and discuss the selection of a suitable evaluation 
method for models used in operational settings in section 6.


\clearpage
%
%
%
\section{2.	Flare Forecasting Tool in Real-Time Operation}
\subsection{2.1.	Procedures of Operational DeFN}

DeFN is designed to predict solar flares occurring in the following 24 h after observing magnetogram images, which are categorized into the two 
categories: ($\geq$M-class and $<$M-class) or ($\geq$C-class and $<$C-class). In the operational system of DeFN forecasts, observation images 
are automatically downloaded, active regions (ARs) are detected, and 79 physics-based features are extracted for each region. Each feature is 
standardized by the average value and standard deviation and is input into the DNN model, DeFN. The output is the flare occurrence probabilities 
for the two categories. Finally, the maximum class of flares occurring in the following 24 h is forecast by taking the maximum probability of the 
forecasts.\\

Operational DeFN was redesigned for automated real-time forecasting with operational redundancy. All the programs written in IDL and Python 
languages are driven by cron scripts at the prescribed forecast issuance time as scheduled. There are a few differences from the original DeFN 
used for research, as explained in the next subsection. A generalized flow chart of operational DeFN is shown in Figure 1.

\clearpage
%
%
\subsection{2.2.	NRT Observation Data}

The first difference between development DeFN and operational DeFN is the use of near real-time (NRT) observation data. We use the observation 
data of line-of-sight magnetograms and vector magnetograms taken from Helioseismic and Magnetic Imager \citep[HMI;][]{sche12, scho12, hoe14} 
on board SDO, ultraviolet (UV) and extreme ultraviolet (EUV) images obtained from Atmospheric Imaging Assembly \citep[AIA;][]{lem12} through 
1600 \,\AA\ and 131 \,\AA\ filters; and the full-disk integrated X-ray emission over the range of 1–8 \,\AA\ observed by GOES. For visualization, 
we also use white light images taken from HMI and EUV images obtained using AIA through 304 \,\AA\ and 193 \,\AA\ filters. The time cadence 
of the vector magnetograms is 12 min, that of the line-of-sight magnetograms is 45 s, those of the 1600 \,\AA\ and 131 \,\AA\ filters are both 
12 s, and that of GOES is less than 1 min.\\

The data product of SDO is provided by the Joint Science Operations Center (JSOC) of Stanford University. The HMI NRT data are generally 
processed and available for transfer within 90 min after observations \citep{lek18}. This is why DeFN was designed to download the observation 
dataset 1 h earlier. If the observation data are unavailable because of processing or transfer delays, the target of downloading is moved back in 
time to 1 to 5 h earlier in the operational DeFN system. When no data can be found beyond 5 h earlier, it is considered that the data are missing. 
Here, the time of 5 h was determined by trial and error. Forecasting takes 20–40 min for each prediction; thus, it is reasonable to set the forecasting 
time to as often as once per hour. The 1 h cadence is comparable to that of the time evolution of the magnetic field configuration in active regions 
due to flux emergence or changes before and after a flare. However, DeFN started operating in the minimum phase of solar activity, so we started 
forecasting with a 6 h cadence instead of a 1 h cadence.\\

The NRT vector magnetograms taken by HMI/SDO are used for operational forecasts, whereas the calibrated HMI `definitive' series of vector 
magnetograms are used for scientific research. The NRT vector magnetograms are accessed from the data series `hmi.bharp\_720s\_nrt' with 
segmentations of `field', `inclination', `azimuth', and `ambig'. These segmentations indicate the components of field strength, inclination angle, azimuth 
angle, and the disambiguation of magnetic field in the photosphere, respectively. Additionally, the NRT line-of-sight magnetograms are downloaded 
from the data series `hmi.M\_720s\_nrt', and the NRT white light images are from the `hmi.Ic\_noLimbDark\_720s\_nrt' (jsoc2) series. The NRT 
data of AIA 131 \,\AA\, 193 \,\AA\, 304 \,\AA\, and 1600 \,\AA\ filters are retrieved from the `aia.lev1\_nrt2' (jsoc2) series.\\

Note that the HMI NRT vector magnetogram is not for the full disk, in contrast to the HMI definitive series data. HMI Active Region Patches (HARP) 
are automatically detected in the pipeline of HMI data processing \citep{bob14}, and the HMI NRT vector magnetogram is limited to the HARP areas 
plus a buffer, on which we overlaid our active region frames detected by DeFN and extracted 79 physics-based features \citep[Figure 2; also see the 
detection algorithms and extracted features in Nishizuka et al. (2017, 2018), and details of the HMI NRT vector magnetogram in][]{lek18}. 
Furthermore, the correlation between the HMI NRT data and the definitive data has not been fully statistically revealed. A future task is to reveal 
how the difference between the HMI NRT and definitive series data affects the forecasting results. The same comments can be made for the AIA 
NRT and definitive series data.\\

\clearpage
%
%
\subsection{2.3.	Implementation of Operational DeFN}

\newcommand{\argmax}{\mathop{\rm argmax}\limits}

Operational DeFN runs autonomously every 6 h by default, forecasting at 03:00, 09:00, 15:00, and 21:00 UT. The forecasting time of 03:00 UT 
was set to be before the daily forecasting meeting of NICT at 14:30 JST. The weights of multi-layer perceptrons of DeFN were trained with the 
2010-2014 observation datasets, and we selected representative hyperparameters by the observation datasets in 2015.\\

For the classification problem, parameters are optimized to minimize the cross entropy loss function. However, since the flare occurrence ratio is 
imbalanced, we adopted a loss function with normalizations of prior distributions. It is the sum of the weighted cross entropy.
\begin{linenomath}
 \begin{equation}
 J_{WCE} = - \sum_{n=1}^N \sum_{k=1}^K w_k y_{nk}^* \log p(y_{nk}).
 \end{equation}
\end{linenomath}
Here, $p(y_{nk}^*)$ is the initial probability of correct labels $y_{nk}^*$, i.e., 1 or 0, whereas $p(y_{nk})$ is the estimated value of probability. The 
components of $y_{nk}^*$ are 1 or 0; thus, $p(y_{nk}^*)$=$y_{nk}^*$. $w_k$ is the weight of each class and is the inverse of the class occurrence 
ratio, i.e., [1, 50] for $\geq$M-class flares and [1, 4] for $\geq$C-class flares. Parameters are stochastically optimized by adaptive moment 
estimation \citep[Adam;][]{kin14} with learning rate = 0.001, $\beta_1$ = 0.9, and $\beta_2$ = 0.999. The batch size was set to 150 \citep[for 
details, see][]{nis18}.\\

Theoretically, the positive and negative events, i.e., whether $\geq$M-class flares occur or not, are predicted in the following manner. The following 
equation is commonly used in machine learning:
\begin{equation}
\hat{y} = \argmax_{k} p(y_k).
\end{equation}
Here, $\hat{y}$ is the prediction result, and the threshold is usually fixed. For example, in the case of two-class classifications, the events with 
a probability greater than 50 \% are output. When we use the model as a probabilistic prediction model, we also tried smaller threshold values 
for safety in operations, although there are no obvious theoretical meanings.\\

Note that the loss function weights cannot be selected arbitrarily. The positive to negative event ratios of $\geq$M-class and $<$M-class 
or $\geq$C-class and $<$C-class flares, which are called the occurrence frequency and the climatological base rate, are 1:50 and 1:4 during 
2010-2015, respectively, in the standard. Only when the cross entropy is applied to the weight of the inverse ratio of positive to negative events 
does it become theoretically valid to output the prediction by equation (2). Therefore, we used the base rate as the weight of cross entropy.\\

The DNN model of the operational DeFN was developed as in Nishizuka et al. (2018). Because the full HMI and AIA datasets obtained from 2010 
to 2015 were too large to save and analyze, the cadence was reduced to 1 h, although in general a larger amount of data is useful for better 
predictions. We divided the feature dataset into two for training and validation with a chronological split: the dataset obtained from 2010 to 2014 
for training and the 2015 dataset for validation. The point of this paper is to contrast how well the DeFN model can predict solar flares in the 
real-time operations and in the research using time series CV methods (=shuffle and divide CV is insufficient). Then, we will discuss that the 
gap between the prediction accuracies in operations and in research using a time-series CV is small (see section 4.3).\\

The time-series CV is stricter than a K-fold CV on data split by active region. It might be true that a K-fold CV on data split by active region 
can also prevent data from a single active region being used in training and testing \citep[e.g.,][]{bob15}. However, a K-fold CV on data split by 
active region allows the training set to contain future samples from different active regions. This may affect the prediction results, when there 
is a long-term variation of solar activity. As well, the number of active regions which produced X-class and M-class flares is not so large that 
a K-fold CV on data split by active region may be biased and not equal.\\

Indeed, solar flare prediction in operation has been done in a very strict condition, where no future data is available. Our focus is not to deny 
a K-fold CV on data split by active region. Instead, our focus is to discuss more appropriate CVs in operational setting.\\ 

The model was evaluated with a skill score, the true skill statistic \citep[TSS;][]{han65}, which is a metric of the discrimination performance. 
Then, the model succeeded in predicting flares with TSS = 0.80 for $\geq$M-class and TSS = 0.63 for $\geq$C-class (Table 1). Note that the 
data for 2016–2018 were not used, because there were fewer flares in this period than in the period between 2010 and 2015.\\

Flare labels were attached to the 2010–2015 feature database for supervised learning. We collected on the disk all the flare samples that occurred 
from the flare event list. We visually checked the locations of the flares, compared them with NOAA numbers, and found the corresponding active 
regions in our database when there were two or more active regions. Then we attached flare labels to predict the maximum class of flares occurring 
in the following 24 h. If $\geq$M-class flares are observed within 24 h after observations, the data are attached with the label (0, 1)$_{\rm M}$; 
otherwise, they are attached with the label (1, 0)$_{\rm M}$. When two M-class flares occur in 24 h, the period with the label (0, 1)$_{\rm M}$ is 
extended. Similarly, the labels (0, 1)$_{\rm C}$ and (1, 0)$_{\rm C}$ are separately attached for the prediction model of C-class flares. The training 
was executed using these labels. On the other hand, in real-time operation, we do not know the true labels of flares, so we attached the NRT 
feature database with dummy labels (1, 0), which are not used in the predictions. It is possible to update the model by retraining it using the latest 
datasets if the prediction accuracy decreases. However, the pretrained operational model is currently fixed and has not been changed.\\
\clearpage
%
%
\section{3. Operation Forecasts Using DeFN}
\subsection{3.1. Graphical Output}

The graphical output is automatically generated and shown on a website (Figure 3). The website was designed to be easy for professional space 
weather forecasters who are often not scientists to understand. Prediction results for both the full-disk and region-by-region images are shown 
on the website, and the risk level is indicated by a mark, ``Danger flares'', ``Warning'', and ``Quiet''. Images are updated every 6 h as new data 
are downloaded. Details of the DeFN website are described below:\\

\begin{itemize}
\item {\bf Solar full-disk images and detected ARs:}\\ 
Images obtained by multiwavelength observations, such as magnetograms and white light, 131 \,\AA\, 193 \,\AA\, 304 \,\AA\, and 1600 \,\AA\ 
images taken by SDO, are shown along with ARs detected by DeFN, where the threshold is set to 140 G in the line-of-sight magnetograms taken 
by HMI/SDO \citep[see details of the detection method in][]{nis17}.\\

\item {\bf Probabilistic forecasts at each AR:}\\
Probabilistic forecasts of flare occurrence at each AR are shown for $\geq$M-class and $<$M-class flares or $\geq$C-class and $<$C-class 
flares by bar graphs, in analogy with the probabilistic forecasts of precipitation. Note that this forecasted probability does not indicate the real 
observation frequency, because the prior distributions are normalized to the peak at 50 \% by the weighted cross entropy, where the loss function 
weights are the inverse of the flare occurrence ratio \citep{nis20}. Thus, operational DeFN is optimized for forecasting with the default probability 
threshold of 50 \%. That is, operational DeFN forecasts flares if the real occurrence probability, which is forecast by the non-weighted cross 
entropy loss function, is greater than the climatological event rate, and it does not forecast flares if the real occurrence probability is less than 
the climatological event rate \citep[see also][]{nis20}. Therefore, the normalized forecasted probability of $\geq$M-class flares sometimes 
becomes larger than that of $\geq$C-class flares.\\

\item {\bf Full-disk probability forecasts and alert marks:}\\ 
The full-disk flare occurrence probability of $\geq$M-class flares, $P_{FD}$, is calculated using
\begin{linenomath}
\begin{equation}
P_{FD} = 1.0 - \prod_{AR_i \in S} (1.0 - P_i),
\end{equation}
\end{linenomath}
where $S$ is the set of ARs on the disk, $i$ is the number of ARs on the disk, element $AR_i$ is a member of set $S$, and $P_i$ is the probabilistic 
forecast at each $AR_i$ \citep{lek18}. The risk level is indicated by a mark based on $P_{FD}$ and is categorized into three categories: ``Danger 
flares'' ($P_{FD}$ $\geq$80 \%), ``Warning'' ($P_{FD}$ $\geq$50 \%), and ``Quiet'' ($P_{FD}$ $<$50 \%). This is analogous to weather forecasting, 
e.g., sunny, cloudy, and rainy.\\

\item {\bf List of comments and remarks:}\\
Forecasted probabilities (percentages), comments, and remarks are summarized in a list.\\

\end{itemize}

\clearpage
%
%
\subsection{3.2. Details of Operations and Redundancy}

Operational DeFN has operated stable forecasts since January 2019. In this subsection, we explain the redundancy and operational details of 
operational DeFN.

\begin{itemize}
\item {\bf Forecast outages:} 
A forecast is the category with the maximum probability of a flare in each of the categories in the following 24 h after the forecast issue time. A 
forecast is normally issued every 6 h. If problems occur when downloading or processing data, the forecast is skipped and handled as a forecast 
failure.

\item {\bf Data outages of SDO/HMI, AIA:} 
There are delays in the HMI/SDO data processing when no applicable NRT data are available for a forecast. In this case, the NRT data to download 
is moved back in time to 1 to 5 h earlier. In such as case, the forecasting target will change from 24 h to 25-29 h, though the operational DeFN is 
not retrained. If no data can be found beyond 5 h earlier, the ``no data'' value is assigned and the forecast is skipped.

\item {\bf No sunspots or ARs with strong magnetic field on disk:} 
If there are no sunspots or ARs detected with the threshold of 140 G on the disk image of the line-of-sight magnetogram, feature extraction is 
skipped, a forecast of ``no flare'' with a probability forecast of 1 \% is issued, and the ``no sunspot'' value is assigned.

\item {\bf Forecasts at ARs near/across limb:} 
DeFN is currently not applicable to limb events. If an AR is detected across a limb, it is ignored in forecast targets.

\item {\bf Flares not assigned to an active region:}  
Detected active regions by operational DeFN are not completely the same as active regions registered by NOAA. There are cases where flares 
occur in decaying or emerging active regions which are not detected by DeFN with the threshold of 140 G. This occurs most often for C-class 
and lower intensity flares, for example, C2.0 flare in NOAA 12741 on 2019 May 15. Such a flare is missed in real-time forecasts but included in 
evaluations.
 
\item {\bf Retraining:} 
DeFN can be retrained on demand, and a newly trained model can be used for forecasting. Currently, the pretrained model is fixed and has not 
been changed so far.

\item {\bf Alternative data input after SDO era (option):} 
Since DeFN is designed to detect ARs and extract features by itself, it can be revised and trained to include other space- and ground-based 
observation data in DeFN, even when SDO data are no longer available.

\end{itemize}

\clearpage
%
%
\section{4. Forecast Results and Evaluation}
\subsection{4.1. Operational Benchmark}

The purpose of machine-learning techniques is to maximize the performance for unseen data. This is called generalization performance. 
Because it is hard to measure generalization performance, it is usually approximated by test-set performance, where there is no overlap 
between the training and test sets.\\

On the other hand, as the community continues to use a fixed test set, on the surface the performance of newly proposed models will seem to 
improve year by year. In reality, generalization performance is not constantly improving, but there will be more models that are effective only 
for the test set. This is partially because that models with lower performance than state-of-the-art are not reported. In other words, there are 
more and more models that are not always valid for unseen datasets. It is essentially indistinguishable whether the improvement is due to an 
improvement in generalization performance or because it is a method that is effective only for the test set.\\

The above facts are well-known in the machine learning community, and the evaluation conditions are mainly divided into two, basic and strict. 
Under the strict evaluation conditions, only an independent evaluation body evaluates each model using the test set only once. The test set is 
not published to the model builders \citep[see e.g.,][]{ard19}. The solar flare prediction is originally a prediction of future solar activity using a 
present observation dataset, and the data available to researchers are only the past data. This fact is consistent with the strict evaluation 
condition in machine-learning community.\\

In this section, we evaluate our operational forecasting results. We call this the ``operational benchmark'' in this paper. In the machine learning 
community, a benchmark using a fixed test set is used only for basic benchmark tests. The basic approach is simple but is known to be insufficient. 
This is because no one can guarantee that the test set is used only once. In strict machine learning benchmarks, evaluation with a completely 
unseen test set is required. Only organizer can see the ``completely unseen test set'', which cannot be seen by each researcher. This is because, 
if researchers use the test set many times, they implicitly tend to select models effective only for the fixed test set.\\

We think that the evaluation methods of operational solar flare prediction models are not limited to evaluations using a fixed test set. However, 
this paper does not deny the performance evaluation using a fixed test set. The purpose of this paper is to show that the operational evaluation 
is important. From a fairness perspective, the strict benchmarking approach takes precedence over the basic approach. Our operational 
evaluation is based on the strict benchmarking approach. We did not retrain our model after the deployment of our system.\\

\clearpage
%
%
\subsection{4.2. Forecast Results and Evaluation}

We evaluated the forecast results from January 2019 to June 2020, when we operated operational DeFN in real time. During this period, 24 C-class 
flares and one M-class flare were observed. The M-class flare was observed on 6 May 2019 as M1.0, which was originally reported as C9.9 and 
corrected to M1.0 later. The forecast results are shown in Table 2. Each contingency table shows the prediction results for $\geq$M-class and 
$\geq$C-class flares. operational DeFN was originally trained with the probability threshold of 50 \% to decide the classification, but in operations, 
users can change it according to their purposes. In Table 2, we show three cases for $\geq$M-class and $\geq$C-class predictions using different 
probability thresholds, such as 50 \%, 45 \%, and 40 \% for reference.\\

Each skill score can be computed from the items shown in contingency tables, and not vice versa. This is a well-known fact. No matter how many 
skill scores you show, you will not have more information than one contingency table. The relative operating characteristic (ROC) curve and the 
reliability diagram, which are shown in Leka et al. (2019), can also be reproduced from the contingency table if it is related to the deterministic 
forecast (forecast of this paper). The ROC curve is a curve or straight line made by plots on a probability of false detection (POFD) - probability 
of detection (POD) plane. The ROC curve for a deterministic forecast is made by connecting three points (0,0), (POFD, POD) for a deterministic 
forecast, and (1,1) \citep[see e.g.,][]{ric00, jol12}. For reference, we introduce skill scores used in Leka et al. (2019), such as the accuracy, Hanssen 
\& Kuiper skill score/Pierce skill score/True skill statistic (TSS/PSS), Appleman skill score (ApSS), equitable threat score (ETS), Brier skill score, 
mean-square-error skill score (MSESS), Gini coefficient, and frequency bias (FB).\\

According to Table 2, the flare occurrence was very rare and imbalanced in the solar minimum phase. Most of the forecasts are true negative. 
When we decrease the probability threshold, the number of forecast events increases. We evaluated our results with the four verification metrics 
in Table 3: accuracy, TSS, false alarm ratio (FAR), and Heidke skill score (HSS) \citep{mur93, bar09, kub17}. They show that operational DeFN 
optimized for $\geq$C-class flare prediction achieved accuracy of 0.99 and TSS of 0.70 with the probability threshold of 50 \%, whereas they 
were 0.98 and 0.83 with the probability threshold of 40 \%. DeFN optimized for $\geq$M-class flare prediction achieved accuracy of 0.99 but TSS 
was only 0.24 because only a single M1.0 flare occurred. Operational DeFN did not predict this flare because it was at the boundary of the two 
categories of $\geq$M-class and $<$M-class flares. This happens a lot in real operations, and this is a weakness of binary classification systems 
used in operational settings.\\

The trends of the contingency tables are similar to those evaluated in the model development phase. (Table 2). However, there are two differences. 
First, the data used were the NRT data, whereas the definitive series was used for development. However, in this case, there was negligible difference 
between them. Second, the evaluation methods are different. The operational DeFN was evaluated on the actual data from 2019 to 2020, whereas 
the development model was validated with the 2010-2014 dataset and tested with the 2015 dataset. It appears that the chronological split provides 
more suitable evaluation results for operations than the common methods, namely, shuffle and split CV and K-fold CV.\\

\clearpage
%
%
\subsection{4.3. Time-series CV}

Here we propose the use of time-series CV for evaluations of operational forecasting models. In previous papers on flare predictions, we used 
hold-out CV, where a subset of the data split chronologically was reserved for validation and testing, rather than the naïve K-fold CV. This is 
because it is necessary to be careful when splitting the time-series data to prevent data leakage \citep{nis18}. To accurately evaluate prediction 
models in an operational setting, we must not use all the data about events that occur chronologically after the events used for training.\\

The time-series CV is illustrated in Figure 4. In this procedure, there are a series of testing datasets, each consisting of a set of observations and 
used for prediction error. The corresponding training dataset consists of observations that occurred prior to the observations that formed the testing 
dataset and is used for parameter tuning. Thus, model testing is not done on data that may have pre-dated the training set. Furthermore, the training 
dataset is divided into training and validation datasets. The model prediction accuracy is calculated by averaging over the testing datasets. This 
procedure is called rolling forecasting origin-based CV \citep{tas00}. In this paper, we call it time-series CV, and it provides an almost unbiased 
estimate of the true error \citep{var06}.\\

Note that the time-series CV has the following advantages: (i) The time-series CV is the standard validation scheme in time-series prediction. 
(ii) A single chronological split does not always reflect low generalization error \citep{bis06}. In other words, the trained model is not guaranteed 
to work for unseen test set. To avoid this, the time-series CV applies multiple chronological splits. The ability to predict new examples correctly 
that differ from those used for training is known as generalization performance \citep{bis06}. Therefore, the time-series CV is more generic and 
appropriate.\\

The evaluation results obtained by time-series CV using the 2010–2017 datasets are summarized in Table 4. The datasets were chronologically 
split to form the training, validation, and testing datasets. TSS is largest with the 2010–2014 datasets for training, the 2015 
datasets for validation, and the 2016 datasets for testing. This is probably because it is not possible to obtain a reliable forecast 
based on a small training dataset obtained from 2010 to 2012. By averaging over the five testing datasets, we found that TSS is 0.70 
for $\geq$M-class flares and 0.59 for $\geq$C-class flares. This procedure will be more suitable for an observation dataset with 
a longer time period.\\

\clearpage
%
%
\section{5. Summary and Discussion}

We developed an operational flare prediction model using DNNs, which was based on a research version of the DeFN model, for operational forecasts. 
It can provide probabilistic forecasts of flares in two categories occurring in the next 24 h from observations: $\geq$M-class and $<$M-class flares 
or $\geq$C-class and $<$C-class flares. DeFN has been continuously used for operational forecasting since January 2019, and we evaluated its 
performance using the forecast and actual flare occurrences between January 2019 and June 2020. We found that operational DeFN achieved an 
accuracy of 0.99 and TSS of 0.70 for $\geq$C-class flare predictions, whereas the accuracy was 0.99 but TSS was only 0.24 for $\geq$M-class flare 
prediction using a probability threshold of 50 \%. using a probability threshold of 40 \%, the accuracy was 0.98 and TSS was 0.83 for $\geq$C-class 
flares, whereas they were 0.98 and 0.48 for $\geq$M-class flares. \\

Operational DeFN has the advantages of a large TSS, good discrimination performance, and the low probability of missed detection of observed flares. 
This is why it is useful for operations that require that no flares are missed, such as human activities in space and critical operations of satellites. 
On the other hand, it tends to over-forecast and the false alarm ratio (FAR) increases. Because the number of true negatives is very large in an 
imbalanced problem such as solar flare prediction, TSS is less sensitive to false positives than to false negatives. Currently, the prior distributions 
of $\geq$M-class and $<$M-class flares are renormalized to increase TSS at threshold probability of 50 \%, but this results in an increase in FAR.\\

When we compared the evaluation results, we observed no significant difference between the pretrained and operational results. This means that, 
at least during January 2019 – June 2020, the difference between NRT and definitive series science data did not greatly affect the forecasts. We 
found a TSS of 0.63 for the $\geq$C-class model evaluated using the pretrained model was maintained and even increased to 0.70 (0.83) for 
operational forecasts with the probability threshold of 50 (40) \%. This suggests that the chronological split is more suitable for the training and 
validation of the operational model than shuffle and split CV.\\

Here, we discuss how to train and evaluate machine learning models for operational forecasting. For an exact comparison, it is desirable to use the 
same datasets among participants. If this is not possible, there are three points that require attention.\\

\begin{enumerate}

\item[(i)] Observation Database: The ratio of positive to negative events should not be artificially changed, and datasets should not be selected 
artificially. Data should be the climatological event rate and kept natural. This is because some metrics are affected by controlling the positive to 
negative event ratio of datasets, especially HSS, which will result in a difference from the operational evaluations. For operational evaluations, it is 
also desirable to include ARs near the limb, although they are excluded in most papers because the values of magnetograms are unreliable owing 
to the projection effect. Currently, in machine learning models, limb flares are not considered, but they also need to be considered in the near future, 
using GOES X-ray statistics as in human forecasting or magnetograms reproduced by STEREO EUV images \citep{kim19}.\\

\item[(ii)] Datasets for Training and Testing: We recommend that a chronological split or time-series CV is used for training and evaluation of 
operational models. Although K-fold CV using random shuffling is common in solar flare predictions, it has a problem for a time-series dataset divided 
into two for training and testing when the time variation is very small, e.g., the time evolution of magnetic field. If the two neighboring datasets, which 
are very similar, are divided into both training and testing sets, the model becomes biased to overpredict flares. It might be true that a K-fold CV 
on data split by active region can also prevent data from a single active region being used in training and testing. However, a K-fold CV on data 
split by active region allows the training set to contain future samples from different active regions. Therefore, in the point of view of generalization 
performance, a time-series CV is stricter and more suitable for operational evaluation.\\

\item[(iii)]	Selection of Metrics: The ranking of models is easily affected by the selection of the metric. Depending on the purpose, users should 
select their preferred model by looking at the contingency tables and skill scores of each model. After understanding that each skill score can 
evaluate one aspect of performance, verification methods should be discussed in the space weather community \citep[see also][]{pag19, cin20}.\\

\end{enumerate}

In this paper, we showed contingency tables of our prediction results. No matter how many skill scores you show, you will not have more information 
than one contingency table. We evaluated our prediction results as a deterministic forecasting model. The ROC curve and the reliability diagram, 
which are shown in Barnes et al. (2016) and Leka et al. (2019), can also be reproduced from the contingency table if it is related to the deterministic 
forecast.\\

We demonstrated the performance of a machine learning model in an operational flare forecasting scenario. The same methods and discussion of 
prediction using machine learning algorithms can be applied to other forecasting models of space weather in the magnetosphere and ionosphere. 
Our future aim is to extend our model to predicting CMEs and social impacts on Earth by extending our database to include geoeffective phenomena 
and technological infrastructures.\\

\clearpage
%
%



\section{Declarations}

\section{Availability of data and materials}

The code is available at https://github.com/komeisugiura/defn18. In the README file, we explain the architecture and selected hyper parameters. 
The feature database of DeFN is available at the world data center of NICT (http://wdc.nict.go.jp/IONO/wdc/). The SDO data are available from 
the SDO data center (https://sdo.gsfc.nasa.gov/data/) and JSOC (https://jsoc.stanford.edu/). The GOES data are available at 
https://services.swpc.noaa.gov/json/goes/.

\section{Competing interests}

The authors declare that they have no competing interests.

\section{Funding}

This work was partially supported by JSPS KAKENHI Grant Number JP18H04451 and NEDO. A part of these research results was 
obtained within ``Promotion of observation and analysis of radio wave propagation'', commissioned research of the Ministry of Internal Affairs and 
Communications, Japan.

\section{Authors' contributions}

N.N., Y.K. and K.S. developed the model. N.N. analyzed the data and wrote the manuscript. M.D. and M.I. participated in discussing the results.

\acknowledgments{We thank all members of JSOC of Stanford University for their support and allowing us to use the SDO NRT data. The data 
used here are courtesy of NASA/SDO, the HMI \& AIA science teams, JSOC of Stanford University, and the GOES team.}




{}






\noindent
%
%
%








\begin{figure}[hbtp]
\centering
\includegraphics[scale=0.55]{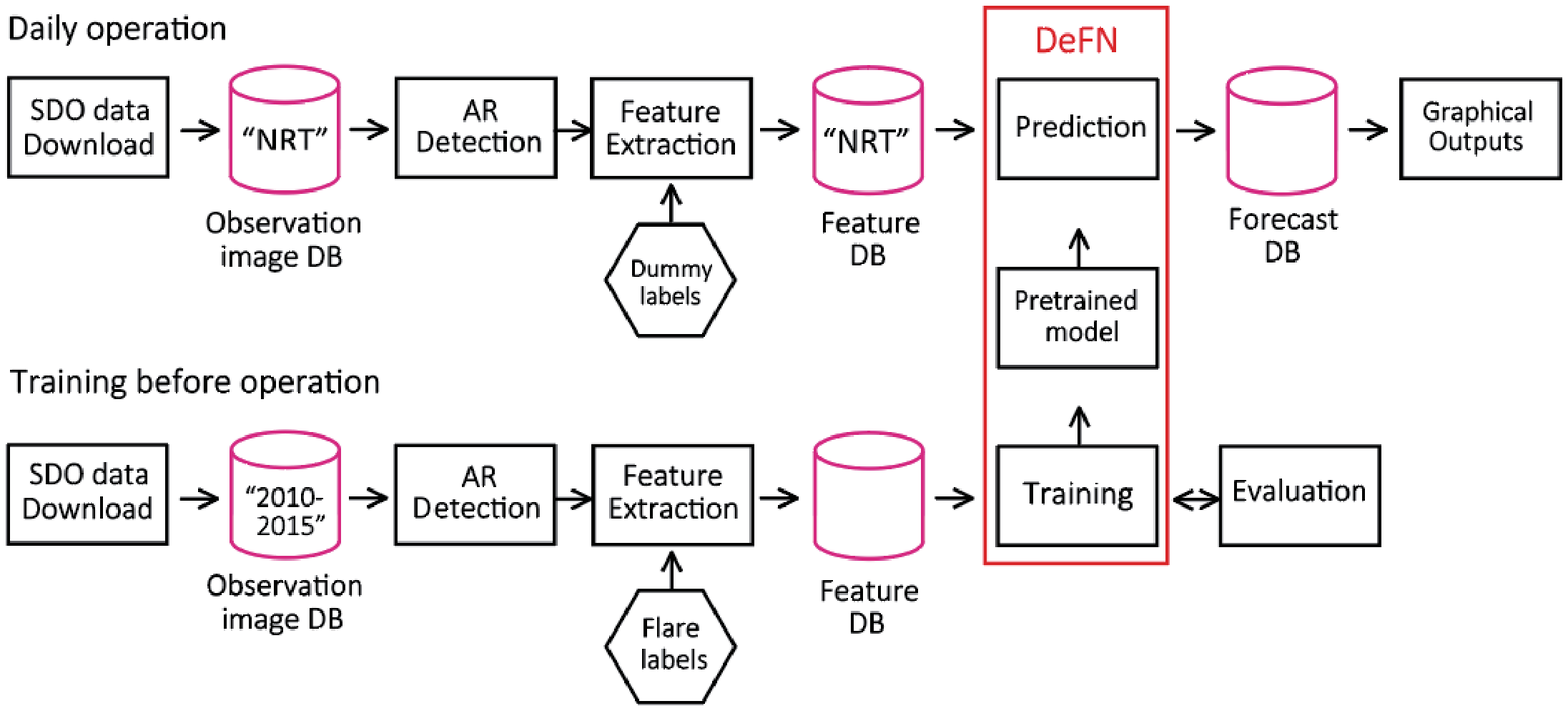}
\caption{Flow chart of operational DeFN. It is executed using the NRT data and the pretrained model. The best-performing model among the models trained using the 2010-2015 datasets is chosen as the pretrained model.\label{fig1}}
\end{figure}

\begin{figure}[hbtp]
\centering
\includegraphics[scale=0.5]{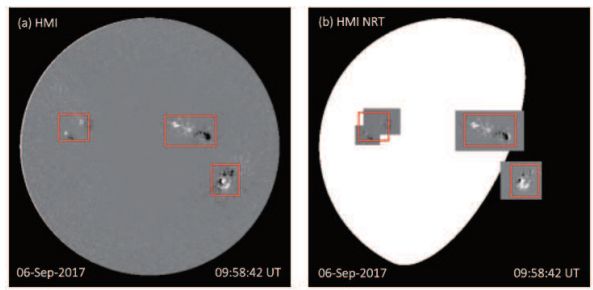}
\caption{Full solar disk vector magnetograms taken by HMI/SDO on 6 September 2017. (a) HMI NRT data showing the HARP areas in gray scale and the areas detected by DeFN with red frames and their numbering. Note that both HARP and DeFN areas overlapped each other. If DeFN areas are outside the HARP areas, the data are set to zero. (b) HMI definitive series data for the full-disk.\label{fig2}}
\end{figure}

\begin{figure}[hbtp]
\centering
\includegraphics[scale=0.6]{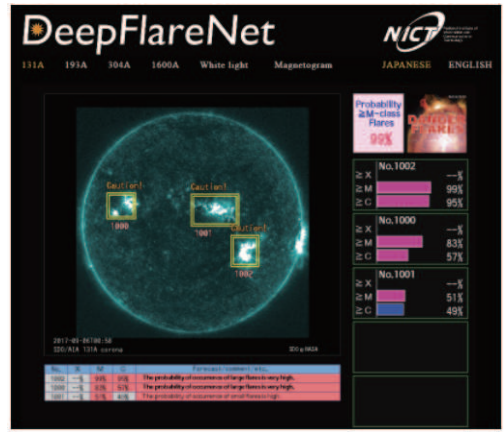}
\caption{Web site of operational DeFN showing a full-disk solar image and detected ARs, in addition to the full-disk forecasts, region-by-region forecasts, an alert mark and a comment list (https://defn.nict.go.jp).\label{fig3}}
\end{figure}

\begin{figure}[hbtp]
\centering
\includegraphics[scale=0.5]{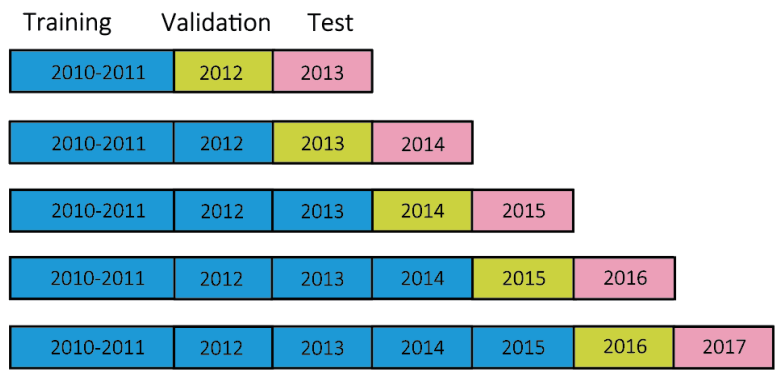}
\caption{Explanation of time-series CV. The series of training, validation, and test sets, where the blue observations form the training sets, the green observations form the 
validation sets, and the red observations form the test sets. \label{fig4}}
\end{figure}

\clearpage
%
%
%
\section{Preparing tables}

%
\begin{table}[h]
\caption{Contingency tables of DeFN using 2010-2015 datasets.}
\begin{minipage}[t]{0.49\textwidth}
\centering
\subcaption{Threshold = 50 \%}
      \begin{tabular}{|c|c||p{1.5cm}|p{1.5cm}|}
      \hline
      \multicolumn{2}{|c|}{$\geq$M-class Flares} & \multicolumn{2}{|c|}{Observed Events} \\
      \cline{3-4}
      \multicolumn{2}{|c|}{ } & Yes & No \\
      \hline\hline
      Forecast & Yes & 963 & 4382 \\
      \cline{2-4}
      Events & No & 54 & 25937 \\
      \hline
      \end{tabular}
\end{minipage}
\begin{minipage}[t]{0.49\textwidth}
\centering
\subcaption{Threshold = 50 \%}
      \begin{tabular}{|c|c||p{1.5cm}|p{1.5cm}|}
      \hline
      \multicolumn{2}{|c|}{$\geq$C-class Flares} & \multicolumn{2}{|c|}{Observed Events} \\
      \cline{3-4}
      \multicolumn{2}{|c|}{ } & Yes & No \\
      \hline\hline
      Forecast & Yes & 4967 & 4420 \\
      \cline{2-4}
      Events & No & 1171 & 20778 \\
      \hline
      \end{tabular}
\end{minipage}
\end{table}

%
\begin{table}[h]
\centering
\caption{Contingency tables of DeFN forecasts in operation from January 2019 to June 2020. They show the forecast results for $\geq$M-class flares and for $\geq$C-class flares, with three different probability thresholds such as 50 \%, 45 \%, and 40 \%.}
\begin{minipage}[t]{0.45\textwidth}
\centering
\subcaption{Threshold = 50 \%}
      \begin{tabular}{|c|c||p{1.5cm}|p{1.5cm}|}
      \hline
      \multicolumn{2}{|c|}{$\geq$M-class Flares} & \multicolumn{2}{|c|}{Observed Events} \\
      \cline{3-4}
      \multicolumn{2}{|c|}{ } & Yes & No \\
      \hline\hline
      Forecast & Yes & 1 & 28 \\
      \cline{2-4}
      Events & No & 3 & 2201 \\
      \hline
      \end{tabular}
\end{minipage}
\begin{minipage}[t]{0.45\textwidth}
\centering
\subcaption{Threshold = 50 \%}
      \begin{tabular}{|c|c||p{1.5cm}|p{1.5cm}|}
      \hline
      \multicolumn{2}{|c|}{$\geq$C-class Flares} & \multicolumn{2}{|c|}{Observed Events} \\
      \cline{3-4}
      \multicolumn{2}{|c|}{ } & Yes & No \\
      \hline\hline
      Forecast & Yes & 27 & 18 \\
      \cline{2-4}
      Events & No & 11 & 2177 \\
      \hline
      \end{tabular}
\end{minipage}
\begin{minipage}[t]{0.45\textwidth}
\centering
\subcaption{Threshold = 45 \%}
      \begin{tabular}{|c|c||p{1.5cm}|p{1.5cm}|}
      \hline
      \multicolumn{2}{|c|}{$\geq$M-class Flares} & \multicolumn{2}{|c|}{Observed Events} \\
      \cline{3-4}
      \multicolumn{2}{|c|}{ } & Yes & No \\
      \hline\hline
      Forecast & Yes & 1 & 31 \\
      \cline{2-4}
      Events & No & 3 & 2198 \\
      \hline
      \end{tabular}
\end{minipage}
\begin{minipage}[t]{0.45\textwidth}
\centering
\subcaption{Threshold = 45 \%}
      \begin{tabular}{|c|c||p{1.5cm}|p{1.5cm}|}
      \hline
      \multicolumn{2}{|c|}{$\geq$C-class Flares} & \multicolumn{2}{|c|}{Observed Events} \\
      \cline{3-4}
      \multicolumn{2}{|c|}{ } & Yes & No \\
      \hline\hline
      Forecast & Yes & 30 & 27 \\
      \cline{2-4}
      Events & No & 8 & 2168 \\
      \hline
      \end{tabular}
\end{minipage}
\begin{minipage}[t]{0.45\textwidth}
\centering
\subcaption{Threshold = 40 \%}
      \begin{tabular}{|c|c||p{1.5cm}|p{1.5cm}|}
      \hline
      \multicolumn{2}{|c|}{$\geq$M-class Flares} & \multicolumn{2}{|c|}{Observed Events} \\
      \cline{3-4}
      \multicolumn{2}{|c|}{ } & Yes & No \\
      \hline\hline
      Forecast & Yes & 2 & 34 \\
      \cline{2-4}
      Events & No & 2 & 2195 \\
      \hline
      \end{tabular}
\end{minipage}
\begin{minipage}[t]{0.45\textwidth}
\centering
\subcaption{Threshold = 40 \%}
      \begin{tabular}{|c|c||p{1.5cm}|p{1.5cm}|}
      \hline
      \multicolumn{2}{|c|}{$\geq$C-class Flares} & \multicolumn{2}{|c|}{Observed Events} \\
      \cline{3-4}
      \multicolumn{2}{|c|}{ } & Yes & No \\
      \hline\hline
      Forecast & Yes & 32 & 34 \\
      \cline{2-4}
      Events & No & 6 & 2161 \\
      \hline
      \end{tabular}
\end{minipage}
\end{table}

%
\begin{table}[h]
\caption{Evaluations of operational forecast results by DeFN from January 2019 to June 2020 with three verification metrics.}
\begin{minipage}[t]{0.9\textwidth}
\centering
\subcaption{$\geq$M-class flare predictions}
      \begin{tabular}{|p{2cm}|p{2cm}|p{2cm}|p{2cm}|p{2cm}|}
      \hline
      Probability threshold & Accuracy & TSS & FAR & HSS \\
      \hline
      50 \% & 0.99 & 0.24 & 0.97 & 0.06 \\
      \hline
      45 \% & 0.98 & 0.24 & 0.97 & 0.05 \\
      \hline
      40 \% & 0.98 & 0.48 & 0.94 & 0.10 \\
      \hline
      \end{tabular}
\end{minipage}
\begin{minipage}[t]{0.9\textwidth}
\centering
\subcaption{$\geq$C-class flare predictions}
      \begin{tabular}{|p{2cm}|p{2cm}|p{2cm}|p{2cm}|p{2cm}|}
      \hline
      Probability threshold & Accuracy & TSS & FAR & HSS \\
      \hline
      50 \% & 0.99 & 0.70 & 0.40 & 0.64 \\
      \hline
      45 \% & 0.98 & 0.78 & 0.47 & 0.62 \\
      \hline
      40 \% & 0.98 & 0.83 & 0.52 & 0.61 \\
      \hline
      \end{tabular}
\end{minipage}
\end{table}

%
\begin{table}
\caption{Evaluation of DeFN forecasts using the time-series CV.}
\begin{tabular}{|c|c|c|} \hline
Datasets & TSS ($\geq$M-class flares) & TSS ($\geq$C-class flares) \\ \hline
Training (2010-2011), Validation (2012), Test (2013) & 0.49 & 0.53 \\ \hline
Training (2010-2012), Validation (2013), Test (2014) & 0.66 & 0.60 \\ \hline
Training (2010-2013), Validation (2014), Test (2015) & 0.77 & 0.66 \\ \hline
Training (2010-2014), Validation (2015), Test (2016) & 0.87 & 0.56 \\ \hline
Training (2010-2015), Validation (2016), Test (2017) & 0.72 & 0.61 \\ \hline
Average & 0.70 & 0.59 \\ \hline
\end{tabular}
\end{table}

\end{document}